\documentclass[onecolumn, amssymb, preprint, showpacs]{revtex4-1}				  

\usepackage{subfigure}   
\usepackage{graphicx}  
\usepackage{dcolumn}			
\usepackage{bm}					
\usepackage{float}			    
\usepackage{amsmath,amsfonts}	
\usepackage{url}					
\usepackage{setspace}		
\usepackage{extarrows}    
\usepackage[symbol]{footmisc}

 \usepackage[hidelinks, colorlinks=true,
linkcolor=blue, citecolor=blue]{hyperref}  
 \usepackage[displaymath, mathlines,right]{lineno}

\makeatletter
\def\makeLineNumberLeft{%
  \linenumberfont\llap{\hb@xt@\linenumberwidth{\LineNumber\hss}\hskip\linenumbersep}
  \hskip\columnwidth
  \rlap{\hskip\linenumbersep\hb@xt@\linenumberwidth{\hss\LineNumber}}\hss}
\leftlinenumbers
\makeatother
\graphicspath{{figure/}}
\usepackage{color}
\newcolumntype{P}[1]{>{\centering\arraybackslash}p{#1}}

\renewcommand\theequation{{\color{blue}\arabic{equation}}}
\begin{document}

\begin{space}      	

\preprint{JASA}		

\title{Achievement of acoustical properties of foam materials by tuning membrane level:\\ Elaborations, Models and Experiments}      

\author{V.H. Trinh\footnote[1] {Also at Le Quy Don Technical University, Hanoi, Vietnam}}			
\author{C. Perrot}			%
\email{camille.perrot@u-pem.fr}
\thanks{Corresponding author.}
\affiliation{Laboratoire Mod\'elisation et Simulation Multi Echelle, MSME UMR 8208 CNRS, Universit\'e Paris-Est, Marne-la-Vall\'ee, France}
\author{V. Langlois}	
\affiliation{Laboratoire G\'eomat\'eriaux et Environnement, LGE EA 4508, 77454, Universit\'e Paris-Est, Marne-la-Vall\'ee, France}
\author{O. Pitois}
\author{Y. Khidas}	
\affiliation{Laboratoire Navier, NAVIER UMR 8205 CNRS, MSME UMR 8208 CNRS, Universit\'e Paris-Est, Marne-la-Vall\'ee, France}



\date{\today}

\begin{abstract}
This work presents a combined numerical and experimental approach to characterize the macroscopic transport and acoustic behavior of foam materials with a membrane cellular structure. A direct link between the sound absorption behavior of a membrane foam-based layer and its local microstructural morphology is also investigated. To this regard, we first produce a set of foam samples having the same density and the same monodisperse pore size but different values of the closure rate of the windows separating the foam pores. Then, the morphology of pore connectivity with membranes is measured directly on SEM together binocular images. The obtained morphological information is used to reconstruct the representative unit cell for computational performance. The knowledge of the computational model of acoustic porous materials obtained by a hybrid approach based on the scaling laws and the semi-phenomenological JCAL model. For validation purpose, the numerical simulations are further compared with the experimental data obtained from a set of three-microphone tube tests, a very good agreement is observed. In acoustic terms, the obtained results point out that for the given high porosity and cell size, we can archive a high sound absorbing ability of based-foam layer by controlling the membrane level at a range of 45-85$\%$. To elaborate these foams, a gelatin concentration in a range of 14-18$\%$ should be used in the foam making process. In addition, we can obtain for instance the peak and the average values of acoustic absorption of a foam layer in a specific frequency range of interest by varying its membrane content. Methodologically, our work proposes (i) a systematic approach to characterize directly macroscopic properties from the local microstructure, and (ii) a manufacturing technique that can be used to make foams with the desired microstructure.
\end{abstract}

\pacs{PACS: XXXX}		

\maketitle




\section{\label{sec1} Introduction}
Noise reduction is of particular concern in many fields, for instance, automotive, aeronautical, and construction industries. Hence, the design task of acoustical materials has recently gained popularity in both academic science and industry. One promising way to solve this task is through the design and optimization of porous structures at some relevant scales. The acoustical macro-behavior of porous materials has been modeled from their microstructure by different methods \cite{gibson1999,perrot2007,kino2008,boeckx2010,doutres2012} for a wide range of porous structures \cite{gasser2005,perrot2008,chevillotte2010,venegas2011,chevillotte2013}. From knowledge about the above micro-macro link, by varying the local geometry of porous materials, several works have focused on defining the optimal microstructure for the target of sound absorption capability \cite{cai2015,lind-nordgren2010,wang1999}.

The rigid-frame porous material assumption is still widely adopted to treat sound absorbing property applications in which the elastic properties of the skeleton do not play a significant role. Two main numerical methods introduced below takes into account only the motions of the fluid phase.

A hybrid method relies on approximate but robust semi-phenomenological model named in the literature as JCAL (Johnson-Champoux-Allard-Lafarge) model \cite{johnson1987,champoux1991,lafarge1997}. This model derives the visco-inertial and  thermal effects from macroscopic parameters with the principle is to solve the local equations governing the asymptotic frequency-dependent visco-thermal dissipation phenomena at the microscopic scale. All the macroscopic parameters of interest can be determined from only three asymptotic calculations (based on the steady Stokes, Laplace, and diffusion-controlled reaction equations) and the frequency-dependent description reconstructed. Contrary, a direct approach solves the linearized Navier-Stokes and the heat equations in harmonic regime in each studied frequency. These parameters allowed use the approximation formulas for the frequency-dependent effective density and effective compressibility \cite{gasser2005,lee2009}. It should be noted that in comparison with the hybrid method the direct approach requires a computational cost by computations of each frequency, especially for cases of complex structure or lager number of computational configurations. For all mentioned above, this work follows up the hybrid framework to model the sound propagation in membrane foams.

Acoustic performance of cellular polymer foams was investigated by various works for both open cell and closed cell structures \cite{gao2016,doutres2013,perrot2012,kino2012,trinh2017}. To this regard, it has been shown that foam membranes, i.e. the solid films closing the windows separating the foam pores, can be of primary importance in acoustical capacity whereas they may occupy a very small volume fraction within the material. Accounting for the membranes effect led to the introduction of dedicated parameters, by measuring simply the fraction of open windows \cite{doutres2012,doutres2013,doutres2014} or distinguishing both fully open and partially open windows \cite{gao2016,yasunaga1996,zhang1999,zhang2012,hoang2012,hoang2014}. From such a refined microstructural description, the homogenization method was found to predict successfully the acoustic properties of several polyurethane foam samples \cite{gao2016,hoang2012,hoang2014}.

However, a complete validation of such method would require considering a set of foam samples allowing for the membrane parameter to be varied within a significant range of values. In this study, we elaborate polymer foam samples showing several membrane contents. The classical numerical homogenization method based on the local geometry models of these real membrane foams is employed to calculate the macroscopic transport properties. The sound absorption performance of foam layers is derived from these values using the JCAL model \cite{johnson1987,champoux1991,lafarge1997}. The numerical results show a good agreement with the measured data using a three-microphone impedance tube. In the end of this work, a systematic micro-macro link is investigated based on our proposed numerical approach. Additionally, some suggestions are also provided to the existing methods, which are used to predict sound absorbing behavior of the partially open cell foams within large range of membrane level.

The remaining part of this paper is organized as follows. Sec. \ref{sec:Method} deals with a brief introduction of the hybrid numerical approach based on the idealized representative unit cell for defining the link between microstructure and properties of acoustic materials. Sec. \ref{sec:Experiment} is devoted the experimental validation involving foam manufacturing process as well as property characterizations of foam sample. In Sec. \ref{sec:Results}, computed results are first further compared with measurements, then some comments in terms of modeling membrane cellular foams are then provided. Finally, in Sec. \ref{sec:conclusion}, we conclude this research and provide directions for future work.

\section{\label{sec:Method} Numerical Method}
\subsection{Equivalent fluid of porous materials}
From the macroscopic perspective, the equivalent-fluid approach is applied where a rigid porous medium is substituted by an effective fluid. This fluid is characterized by the effective density \cite{johnson1987} and effective bulk modulus \cite{champoux1991, lafarge1997} as follows:
\begin{equation}
\rho(\omega)=\frac{\rho_0}{\phi}\left[\alpha_{\infty}+\frac{\phi\sigma}{j\omega\rho_0}\sqrt{1+j\omega\frac{\rho_0}{\eta}\left(\frac{2\eta\alpha_{\infty}}{\sigma\phi\Lambda}\right)^2}\right]
\end{equation}
and
\begin{equation}
K(\omega)=\frac{\gamma{P_0}/\phi}{\gamma-(\gamma-1)\left[1-j\frac{\phi\kappa}{k'_0C_p\rho_0\omega}\sqrt{1+j\frac{4k'_0 P_r\rho_0\omega}{\kappa\Lambda'^2\phi^2}} \right]^{-1}}
\end{equation}
in which $\rho_0$ and $\eta$ denote the density and dynamic viscosity of the ambient fluid (i.e. air).  $\kappa=\gamma{P_0}$ is the air adiabatic bulk modulus, $P_0$ the atmospheric pressure, and $\gamma=C_p/C_v$ is the ratio of heat capacity at constant pressure ($C_p$) to the heat capacity at constant volume ($C_v$).

The JCAL effective fluid model involves 6 macroscopic parameters ($\phi,~\Lambda',~\sigma$,~$\alpha_{\infty},~\Lambda,~ k'_0$) in order to desribe visco-inertial as well as thermal dissipative effects inside the porous media. In which, the porosity $\phi$ and thermal characteristic length $\Lambda$ are defined directly from the local geometry, and others are computed from numerical solutions of (i) the Stoke equations \cite{johnson1987} (the static air flow resistivity $\sigma$); (ii) the inertial equations \cite{johnson1987} (the high frequency tortousity $\alpha_{\infty}$ and the viscous characteristic length $\Lambda'$) and (iii) the diffusion equations \cite{rubinstein1988} (for the static permeability $k'_0$), (detailed view see \cite{perrot2012} and Appendix A).

In acoustic terms, a homogeneous layer is described by the wave number $k_c(\omega)$ and the characteristic impedance $Z_c(\omega)$ as follow \cite{allard2009},
 \begin{equation}
 k_c(\omega)=\omega\sqrt{\rho(\omega)/K(\omega)},\ \
 Z_c(\omega)=\sqrt{\rho(\omega)K(\omega))}
 \end{equation}
 The normal incidence sound absorption coefficient of this porous layer is derived from the complex reflection factor by,
 \begin{equation}
 \alpha=1-\Bigg|\frac{Z_s(\omega)-Z_0}{Z_s(\omega)+Z_0}\Bigg|^2
 \end{equation}
 with $Z_0$ is the air impedance and $Z_s(\omega)$ is the normal incidence surface impedance. For a layer of thickness $L_s$, $Z_s(\omega)$ is given as, $Z_s(\omega)=-jZ_c(\omega)\text{cot}(k_c(\omega)L_s)$.
\subsection{Representative unit cell of foams}
The idealized Kelvin's tetrakaidecahedron is widely used for modeling high porous foams \cite{weaire2001}. This space-filling arrangement of idelical cell is a good presentative structure for real cellular foams with equal-sized bubbles or cells of equal volume \cite{perrot2007}. The cross section of struts of this framework is modeled in different shapes such as circle, triangular or concave shape, interestingly, the ligament shape has limited influence on macroscopic acoustic properties \cite{doutres2013,perrot2012}, it means that we can able to treat this shape in simple one (i.e. triangular). A periodic unit cell (PUC) is used to represent the local structure of our foam samples (see an unit cell (Fig. (\ref{fig:FIG2}.e))  and the corresponding finite element mesh (Fig. \ref{fig:FIG2}(f)). The cell is based on an order packing of 14-sided polyhedron with 6 squared faces and 8 hexagonal faces. As we are mostly interested in the effect of the closure rate of windows, the cell skeleton is made of idealized ligament having a length   and an equilateral triangular cross section of edge side  (see Appendix D). For PUC in a case of membrane closed cell structure, the morphology of films in window faces of material framework will be characterized corresponding to local property of membrane in the next section (for more details, see Sec. \ref{subsec:charac} and Supplemental material).
\section{\label{sec:Experiment} Experimental Validation}
\subsection{Elaboration of controlled polymer foams}
We elaborate solid polymer foam samples having a gas volume fraction $\phi$  and a monodisperse pore size $D_b$, but a tunable membrane content. The foam making procedure can be described as follows (see Fig. \ref{fig:FIG1}):

 (1) Monodisperse precursor aqueous foam is generated. Foaming liquid, i.e. TTAB at 3 g/L in water, and nitrogen are pushed through a T-junction allowing controlling the bubble size by adjusting the flow rate of each fluid. Produced bubbles are stored in a glass column and a constant liquid fraction over the foam column is set at 0.99 by imbibition from the top with foaming solution.

  (2) An aqueous gelatin solution is prepared at a mass concentration $C_{gel}$ within the range 12-18$\%$. The temperature of this solution is maintained at T $\approx\ 60^\circ$C in order to remain above the sol/gel transition ($30^\circ$C).

   (3) The precursor foam and the hot gelatin solution are mixed in a continuous process thanks to a mixing device based on flow-focusing method \cite{haffner2015}. By tuning the flow rates of both the foam and the solution during the mixing step, the gas volume fraction can be set,  = 0.8. Note also that the bubble size is conserved during the mixing step. The resulting foamy gelatin is continuously poured into a cylindrical cell (diameter: 40 mm and height: 40 mm) which is rotating around its axis of symmetry at approximately 50 rpm. This process allows for gravity effects to be compensated until the temperature decreases below the setting temperature.

    (4) The cell is let one hour at $0^\circ$C, then one week in a climatic chamber (T = $20^\circ$C and RH = 30$\%$). During that stay, water evaporates from the samples and the gas fraction increases significantly.

     (5) After unmolding, a slice (thickness: 20 mm and diameter: 40 mm) is cut (see Fig. \ref{fig:FIG3}(c).

\begin{figure}[!h]
\centering \includegraphics[width=0.7\textwidth, trim={0cm 0cm 1cm 0cm},clip]{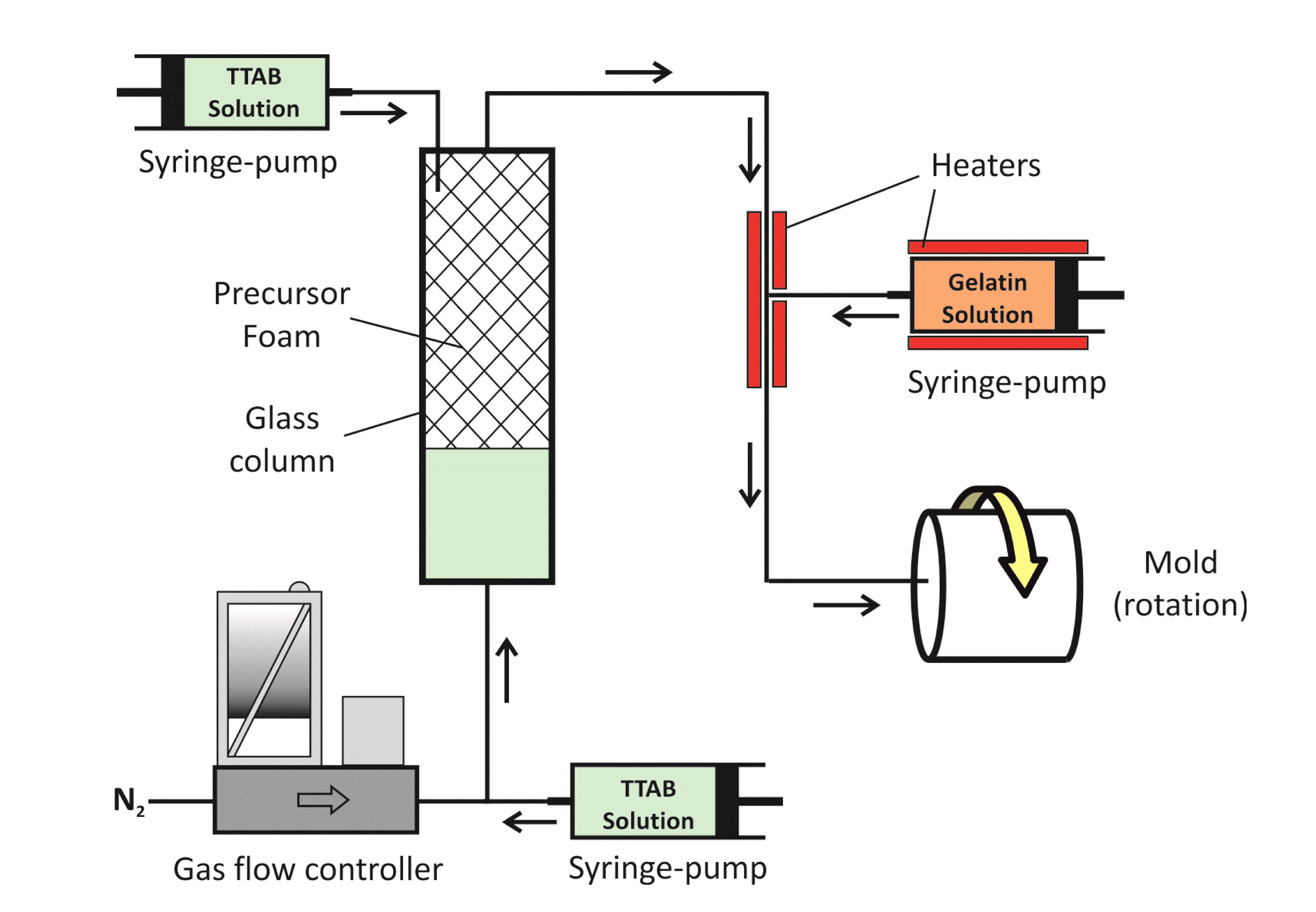}
\caption{\label{fig:FIG1}{(Color online) Diagram of foam making process}}
\end{figure}

\subsection{\label{subsec:charac}Characterization of the foam samples}
As a density of the dried gelatin was measured to be 1.36 kg/m$^3$, the volume and weight measurements of the prepared samples give values of pore volume fraction. For the gelatin concentrations used in this study, the pore volume fraction was found to vary between 0.977 and 0.983, so that in the following we will consider that this parameter is approximately constant and equal to  $0.980\pm0.003$. Observation of the cylindrical surface of the sample (see Fig. \ref{fig:FIG2}(a) allows for the pore (bubble) size to be measured:  = 810 $\mu$m (the absolute error on   is $\pm30\ \mu$m) for all samples. In addition, the shape anisotropy degree was estimated through a ratio as   (see Fig. \ref{fig:FIG2}(b)), note that this degree is also considered in both axial and radial directions), providing values smaller than 1.15 for all samples, which justify to neglect this effect in the following. The membrane content is evaluated by measuring the closure rate of windows separating the pores. We proceed as follows: over several hundred windows observed on both the top and bottom sample surfaces, the proportion of fully closed windows $x_{fc}$ is measured. For the partially closed windows $x_p$, with a proportion equal to (1-$x_{fc}$), their average closure degree is also measured: $r_c=1-\sqrt {A_{elip}/A_{poly}}$  , where  $A_{poly}$ is the window area (the area of the corresponding polygonal face) and  $A_{elip}$ is the aperture area (the area of the fitting ellipse with the aperture) (see Fig. \ref{fig:FIG2}(c)). According to these notations, the membrane closed fraction writes: Note that in order to get all the structural information required for the modeling (for PUC mentioned previously), we refine this treatment by distinguishing the windows counting 4 or less edges (referred to as $'sq'$), from the windows counting more than 4 edges (referred to as $'hex'$). The global closure rate of the cell can be tuned by varying the number of partially closed windows, i.e.   and  as well as the closure level of those windows, i.e. $r^{sq}$ and $r^{hex}$ respectively. The number of fully closed windows is equal to: $N_{fc}^{sq}=6- N_p^{sq},\ N_{fc}^{hex}=8-N_p^{hex}$.

The structural characterization was completed by a measurement of the membrane thickness through SEM microscopy (Fig. \ref{fig:FIG2}(d)). From ten SEM images we obtain an average thickness equal to 1.5$\pm0.25\ \mu$m, which is close to thicknesses measured for similar polymer foams \cite{gao2016,yasunaga1996,hoang2012}.
\begin{figure}[!h]
\centering \includegraphics[width=0.9\textwidth, trim={0cm 0cm 0cm 0cm},clip]{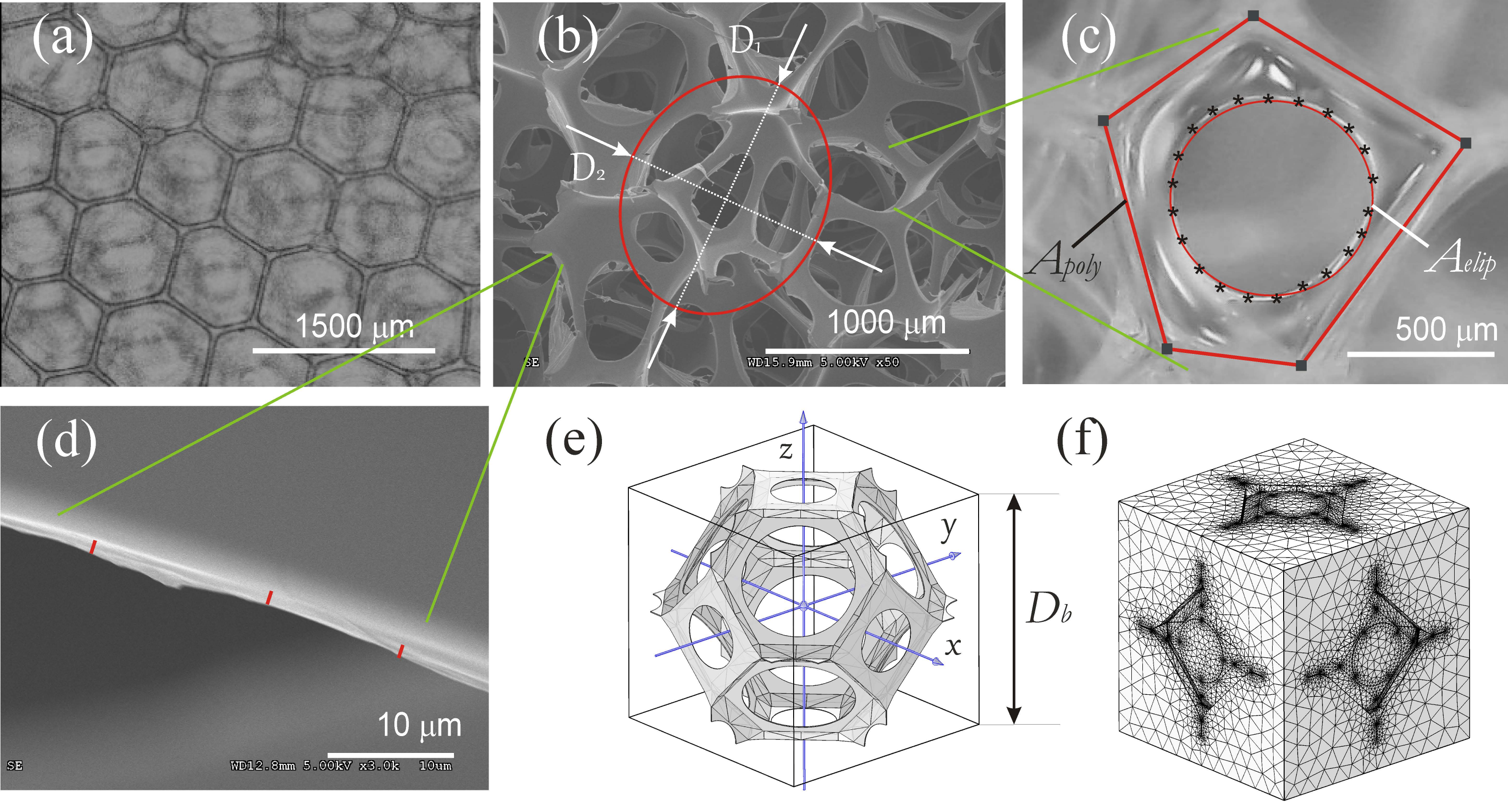}
\caption{\label{fig:FIG2}{(Color online) Characterizations of dried-gelatin foams materials: top view of foam sample (a), anisotropy degree (b, membrane closure rate (c), and membrane thickness (d) measurements, an unit cell (e) and its finite element mesh (f).}}
\end{figure}

The static resistivity $\sigma$ of foam samples is obtained from the measured differential pressures $\Delta p$ and the controlled steady laminar flow rate $Q$ \cite{stinson1988}, according to the standard ISO 9053 (method A): $\sigma=\Delta p A_s/QL_s$ with $A_s$ and $L_s$ are the cross-section area and thickness of sample, $A_s$=12.57 cm$^2$, $L_s$=20 mm. The relative error of this measure is lower than 10$\%$.

Acoustic properties are determined with a 3-microphone impedance tube \cite{iwase1998,salissou2010} (length: 1 m, diameter: 40 mm) (see Fig. \ref{fig:FIG3}(a). Separating distances are as: Micro. $\#$1-Micro. $\#$2: $d_{12}$=35 mm, Micro. $\#$2-Sample: $d_{2s}$=80 mm, and Sample-Micro. $\#$3: 0 mm (Fig. \ref{fig:FIG3}(b)). It is kept in mind that the diameter  of the samples was slightly larger than 40 mm so that air leakage issue and sample vibration were successfully avoided. The test frequency  ranges from 4 Hz to 4500 Hz with a step size of 4 Hz.
\begin{figure}[!h]
\centering \includegraphics[width=0.7\textwidth, trim={0cm 0cm 0cm 0cm},clip]{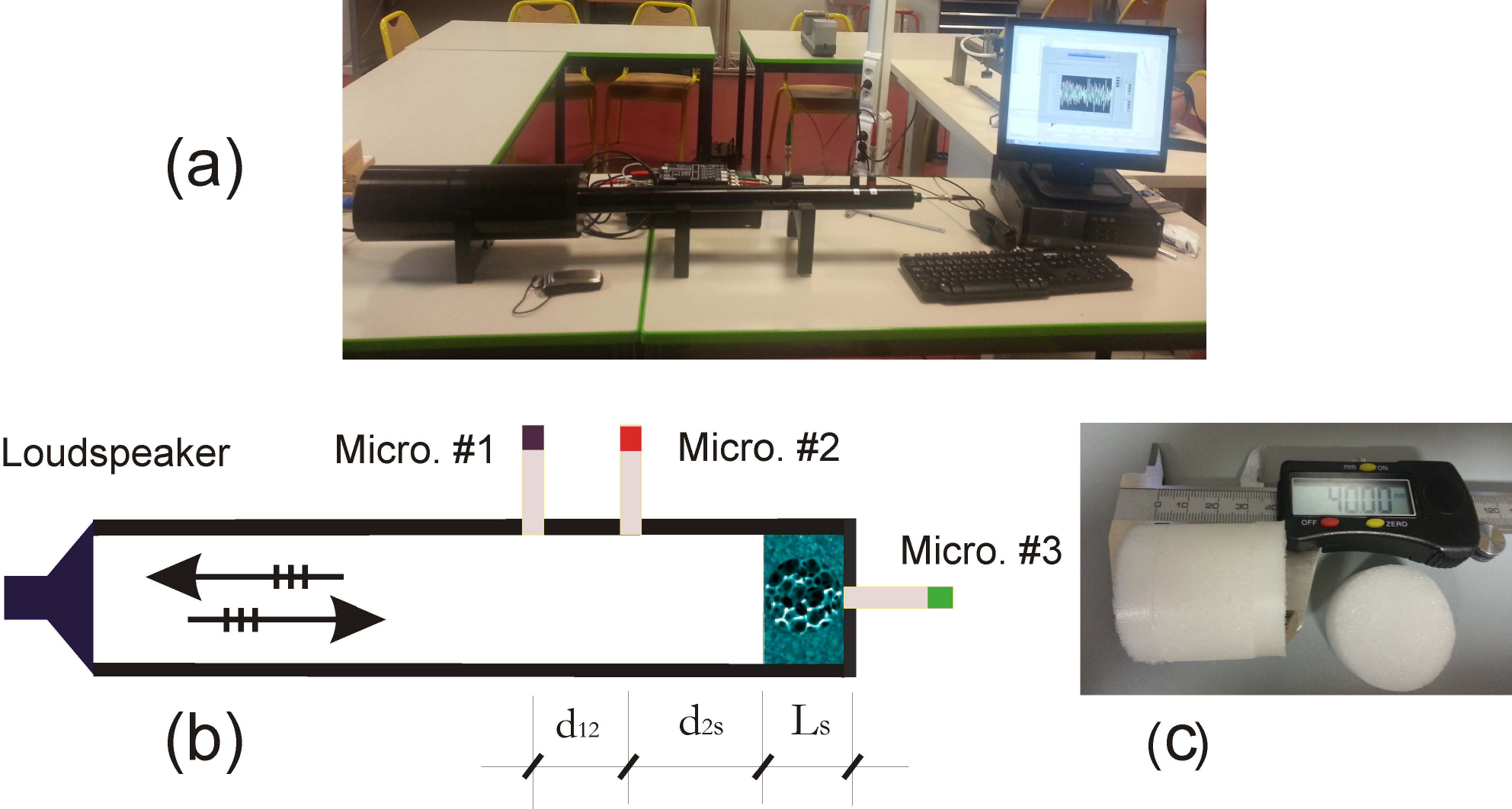}
\caption{\label{fig:FIG3}{(Color online) Three-microphone impedance tube measurement: experimental set-up (a), tube macroscopic configuration (b), and the foam sample (c).}}
\end{figure}

\section{\label{sec:Results} Results and Discussion}
\subsection{Non-acoustic property}
As presented in Table \ref{tab:t1}, it can be revealed that the gelatin concentration $C_{gel}$ (varying from 12$\%$ to 18$\%$) in the foaming solution controls the membrane fraction and influences on structural characterization.
\begin{figure}[!h]
\centering \includegraphics[width=0.75\textwidth, trim={0cm 0cm 0cm 0cm},clip]{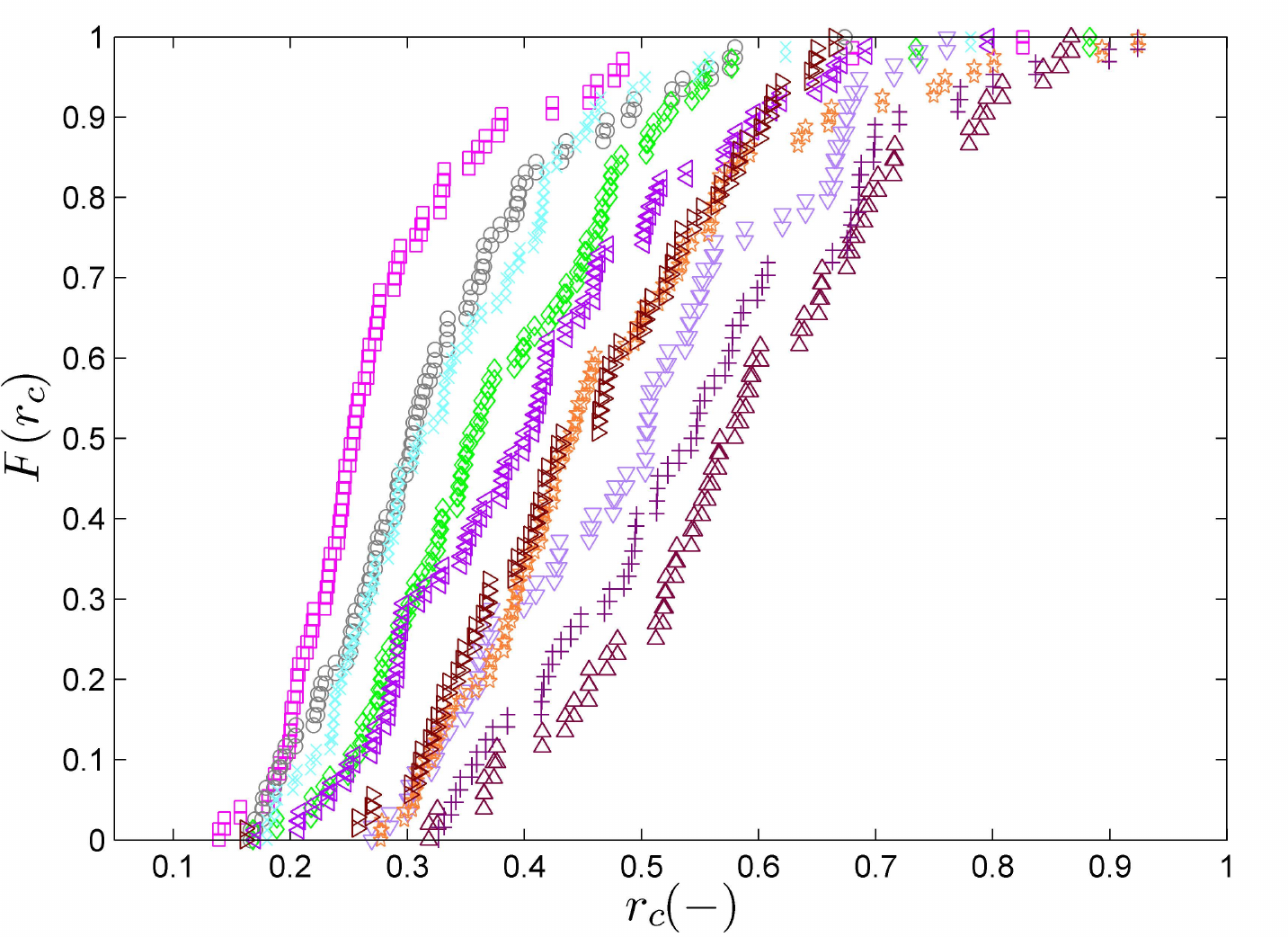}
\caption{\label{fig:FIG4}{(Color online) Distribution function of membrane closure rate measured on foam samples (D20 ($\Box$), D14 ($\circ$), D1($\Diamond$), D2($\times$), B9($\bigtriangledown$), B16($\lhd$),  B5($\star$), B4($\rhd$), B2($\bigtriangleup$), and B1($+$)).}}
\end{figure}
\begin{table}[h]
\centering
\caption{Static airflow resistivity and microstructural parameters measured on samples}
\begin{tabular*}{1\columnwidth}{@{\extracolsep{\fill}}lcccccccccc}
\hline
\hline
\noalign{\smallskip}\noalign{\smallskip}
Foam&$C_{gel}$
&$x^{sq}_{p}$&$x^{sq}_{fc}$
&$x^{hex}_{p}$&$x^{hex}_{fc}$&$r_c$
&$N^{sq}_{fc}$&$N^{hex}_{fc}$&$f_c$\\
\noalign{\smallskip}
&$(\%)$&$(\%)$
&$(\%)$&$(\%)$&$(\%)$
&$(-)$&$(-)$&$(-)$&$(\%)$\\
\noalign{\smallskip}
\hline
\noalign{\smallskip}
D20&12 &20.8 & 4.2 &72.1 & 2.6 &0.280 &1 &0 & 33.1\\
D14&13 &18.4 & 8.1 &64.9 & 8.6 &0.323 &2 &1 & 43.6\\
D1&16  &15.0 &12.4 &64.0 & 8.6 &0.334 &3 &1 & 47.4\\
D2&16  &12.7 &16.1 &56.6 &14.6 &0.397 &3 &2 & 57.0\\
B9&16  &10.4 &19.8 &49.2 &20.6 &0.409 &4 &2 & 64.8\\
B16&17 & 8.2 &23.6 &45.5 &22.7 &0.495 &4 &3 & 72.9\\
B5&18  & 6.6 &22.1 &42.6 &28.7 &0.473 &5 &3 & 74.1\\
B4&18  & 6.1 &17.9 &26.1 &50.9 &0.448 &4 &5 & 83.2\\
B2&18  & 6.3 &18.2 &15.8 &59.7 &0.584 &4 &6 & 90.8\\
B1&18  & 3.4 &24.9 &11.2 &61.3 &0.551 &5 &7 & 94.2\\
\noalign{\smallskip}
\hline
\end{tabular*}
\label{tab:t1}
\end{table}
\begin{figure}[!h]
\centering \includegraphics[width=1\textwidth, trim={0cm 0cm 1cm 0cm},clip]{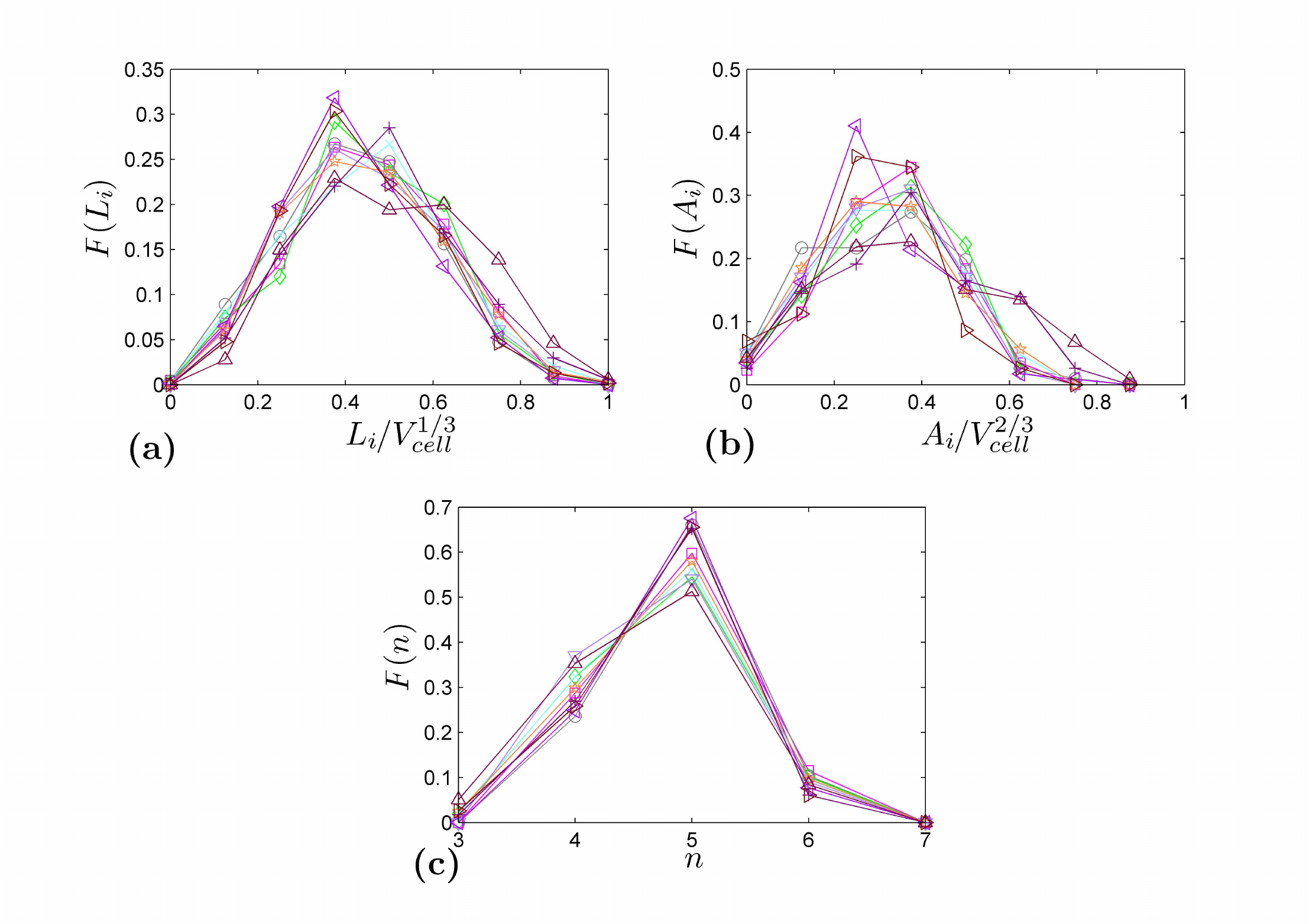}
\caption{\label{fig:FIG5}{(Color online) Morphological properties measured on foam samples (D20 ($\Box$), D14 ($\circ$), D1($\Diamond$), D2($\times$), B9($\bigtriangledown$), B16($\lhd$),  B5($\star$), B4($\rhd$), B2($\bigtriangleup$), and B1($+$)): normalized edge length distributions(a), normalized face area distributions (b), and distributions of faces with $n$ edges (c). Note that $V_{cell}$ is the unit cell volume of $D_b^3/2$}.}
\end{figure}
For all samples, the ratio $(x_{fc}^{sq}+x_p^{sq})/(x_{fc}^{hex}+x_p^{hex})$ is close to $1/3$, which is consistent with previous works \cite{matzke1946,gong2005}. Details in properties of membrane morphology, the closure rate of membrane in the larger windows $r_c^{hex}$  tends to be slightly smaller than that in small ones $r_c^{sq}$ for all sample foams. Consequently, this allows to consider the PUCs has an identical closure rate $r_c=(r_c^{sq} x_p^{sq}+r_c^{hex} x_p^{hex})/(x_p^{sq}+x_p^{hex})$  instead of the identical membrane size mentioned in \cite{hoang2012}. In addition, the measured morphological properties shown in Fig. \ref{fig:FIG5} are very close with these provided in Refs \cite{kraynik2003, koll2014} for foams with monodisperse or relaxed structure.

 As shown in Fig. \ref{fig:FIG4}, it is worth noting that the level of closure rate $r_c$ increases upon to the rising of proportion of fully closed windows $x_{fc}$, this leads to an increase of membrane content from foam D20 to foam B1 (as shown in the last column in Table \ref{tab:t1}. Based the open cell unit shown before and these morphological information, several type of PUC are reconstructed (see Fig. 4). It can be seen that each PUC has different set of parameters: $N_{fc}^{sq},\ N_p^{sq},\ N_{fc}^{hex},\ N_p^{hex}$ and the closed rate $r_c$. It should be noted that airflow resistivity corresponding to samples are characterized value obtained from the imaginary part of the low frequency behavior of the effective density as expression \cite{panneton2006,only2008} (see Eq. (\ref{eq:charac-resis}) in Appendix D).
The effective density is assumed from the previous three-microphone impedance tube. The measured results of static resistivity show a significant increase upon the corresponding foam samples owning the membrane level of $f_c$ from 30$\%$ to 90$\%$ (see permeablity results shown Fig. \ref{fig:FIG6} and Table S.VI in Supplemental Materials). This trend is very consistent with experimental data \cite{perrot2012,kino2012,hoang2014} and semi-empirical result provided by works from Doutres et al. \cite{doutres2013,doutres2011} (see also in Eq. (\ref{eq:semi-resis1}) in Appendix C).

In terms of numerical results of non-acoustic properties, the configuration of each unit cell is considered based on its distribution of the fully closed faces (see also Supplemental Material). Of course, this distribution has no influence on the thermal characteristic length $\Lambda'$ and the porosity $\phi$. Other parameters are computed based on an averaging conductivity as follows. As in reconstruction of the representative cell, the PUC involves $N_{fc}^{sq}$ of fully closed squares and $N_{fc}^{hex}$ of fully closed hexagons. It could be easy to define the number of possible configurations $n_{cf}$ as the following expression:
\begin{equation}
n_{cf}=	\dbinom{6}{N_{fc}^{sq}}\times \dbinom{8}{N_{fc}^{hex}}
\end{equation}
It is noted that the unit cell is not fully symmetry, so the averaging macroscopic transport property of each configuration $\bar{\tau}^i$ over all the directions is calculated as \cite{malinouskaya2009},
\begin{equation}
\bar{\tau}^i=\frac{1}{3}\text{tr}(\boldsymbol{\tau}^i)
\end{equation}
In which, the ${\tau_{xx}^i},\ {\tau_{yy}^i},{\tau_{zz}^i}$ are three components of conductivity $\boldsymbol{\tau}^i$ along the $x$, $y$ and $z$ direction, respectively. Noted that $n_{cf}$ is total possible configurations of the PUC, so $\boldsymbol{\tau}_{xx}^i\equiv \boldsymbol{\tau}_{yy}^i \equiv \boldsymbol{\tau}_{zz}^i$, with $\boldsymbol{\tau}_{kk}^i=\{{\tau}_{kk}^i \mid i=1,2,..,n_{cf}\}$ and $kk=xx,~yy,~zz$. In addition, we are able to define the conductivity along $y$ and $z$ direction of a configuration from that along $x$ direction of other configurations (Detail about defining these configurations, see Supplemental Material).

Then, the averaging macroscopic conductivity is deduced by,
\begin{equation}
\bar{\tau}=\frac{1}{n_{cf}}\sum_{i=1}^{n_{cf}}\bar{\tau}^i=\frac{1}{n_{cf}}\sum_{i=1}^{n_{cf}}{\tau_{kk}}^i
\end{equation}
As shown in Fig. \ref{fig:FIG6}, the computed and the characterized non-dimensional macroscopic parameters are in good agreement in term of their dependence on the membrane fraction $f_c$. In general, the membrane content has a significant effect on all macroscopic transport properties, and these influences are in good agreement with numerical results \cite{hoang2012} and imperial equations \cite{doutres2013}, note that our parameter $f_c$ has a directly link with both the membrane closure rate and the reticulated rate in these reference works (see Sec. \ref{subsec:Results-2}. Following an increase of $f_c$, two characteristic lengths $(\Lambda',\ \Lambda)$ and viscous permeability $k_0$ decrease sharply, while high frequency tortuosity $\alpha_{\infty}$ shows a significant increase. This leads to an interesting sound absorption performance of membrane foam layer indicated in the following section.
\subsection{\label{subsec:Results-2} Acoustic property}
In terms of sound absorption ability, as illustrated in Fig. \ref{fig:FIG7}, the results show that this class of foam samples has distinct behaviors in term of sound performance for given frequency range, also our simulated results are in agreement with experimental data and characterized results. Generally, it is clearly that the membrane fraction of foams has strongly influence on their sound absorption performance in different frequency bands in the range from 4 to 4500 Hz.
The samples owning the lowest (foam D20) and highest (foams B2 ($\alpha\approx0.2$) and B1 ($\alpha\approx$0.15) membrane fraction show a poor sound absorbing capacity ($\alpha \le 0.5)$ in whole range of frequency (for the sample B1, see Fig. S.9 in Supplemental Material). It means that in this scale of cell size, a layer made of opened-cell or closed-cell foams is not good for the sound absorption applications. Of course it is here not mentioned the case that its absorbing coefficient is improved by increase in its thickness. Interestingly, with the middle level of membrane content, the absorption property is significantly improved. In the trend of increasing of $f_c$ (sample D1, D2, and B9), as mentioned previously, the enhanced low frequency sound absorption correlates with an increase in tortuosity with a decrease of the other transport parameters. For samples (B16, B4 and B5), however, this enhanced of mean value of SAC is accompanied with a decrease of the peak of SAC in the whole frequency range.
Fig. \ref{fig:FIG8} illustrates of a sound absorbing chart depended on frequency together membrane levels, it is seen that the computational chart (the left part) is very close with the experimental one (the right part). These charts figured out clearly that the range of membrane level produces a high sound absorption ability in a large frequency range of interested. In addition, the charts also demonstrate the moving of location of frequency with a SAC peak and the high SAC in the low frequency range of high membrane foams.

\begin{figure}[!h]
\centering \includegraphics[width=1\textwidth, trim={0cm 0cm 1cm 0cm},clip]{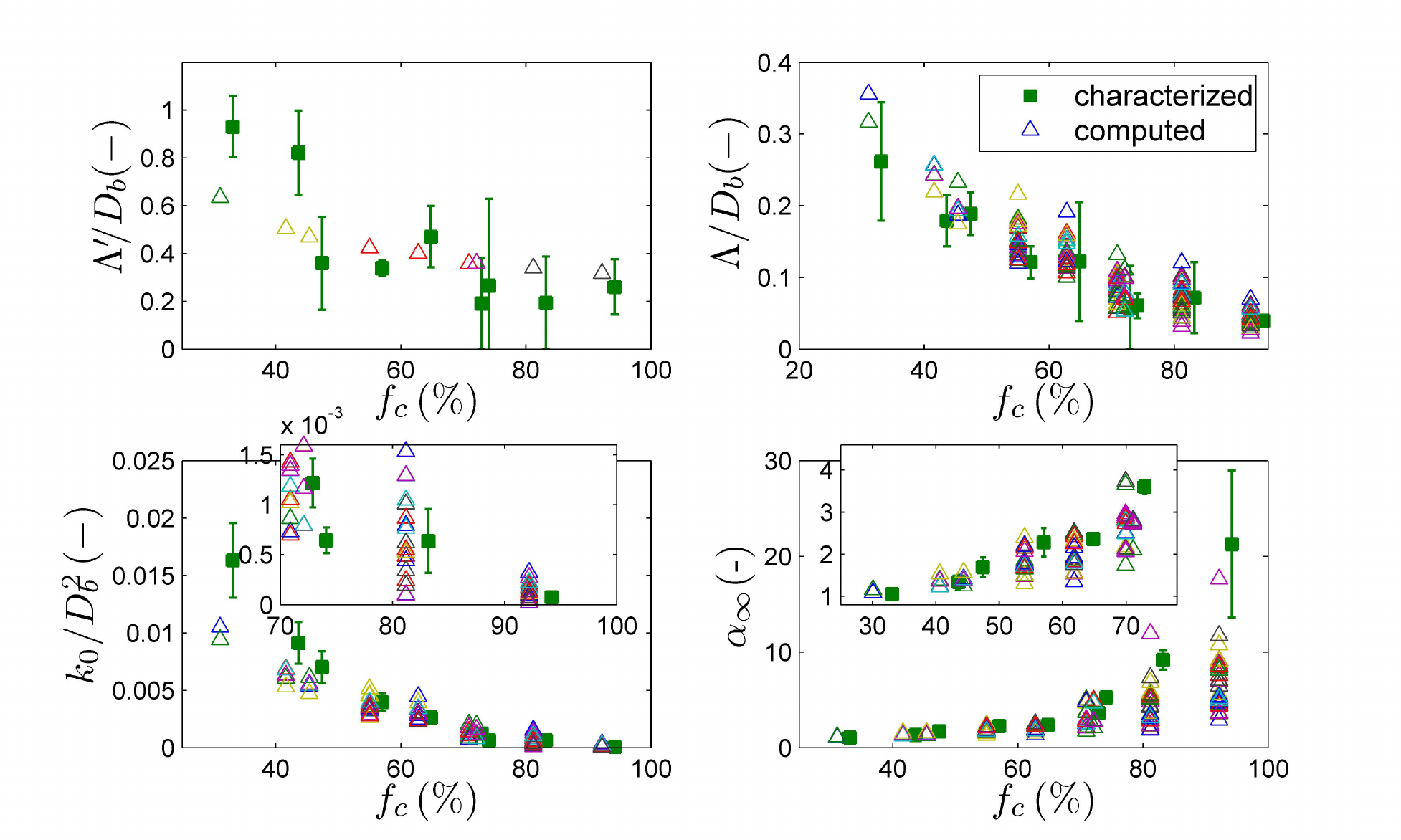}
\caption{\label{fig:FIG6}{(Color online) Dependence of  characterized($\Box$) and computated ($\triangle$) dimensionless transport properties on the membrane fraction.}}
\end{figure}

\begin{figure}[!h]
\centering \includegraphics[width=1\textwidth, trim={0cm 0cm 1cm 0cm},clip]{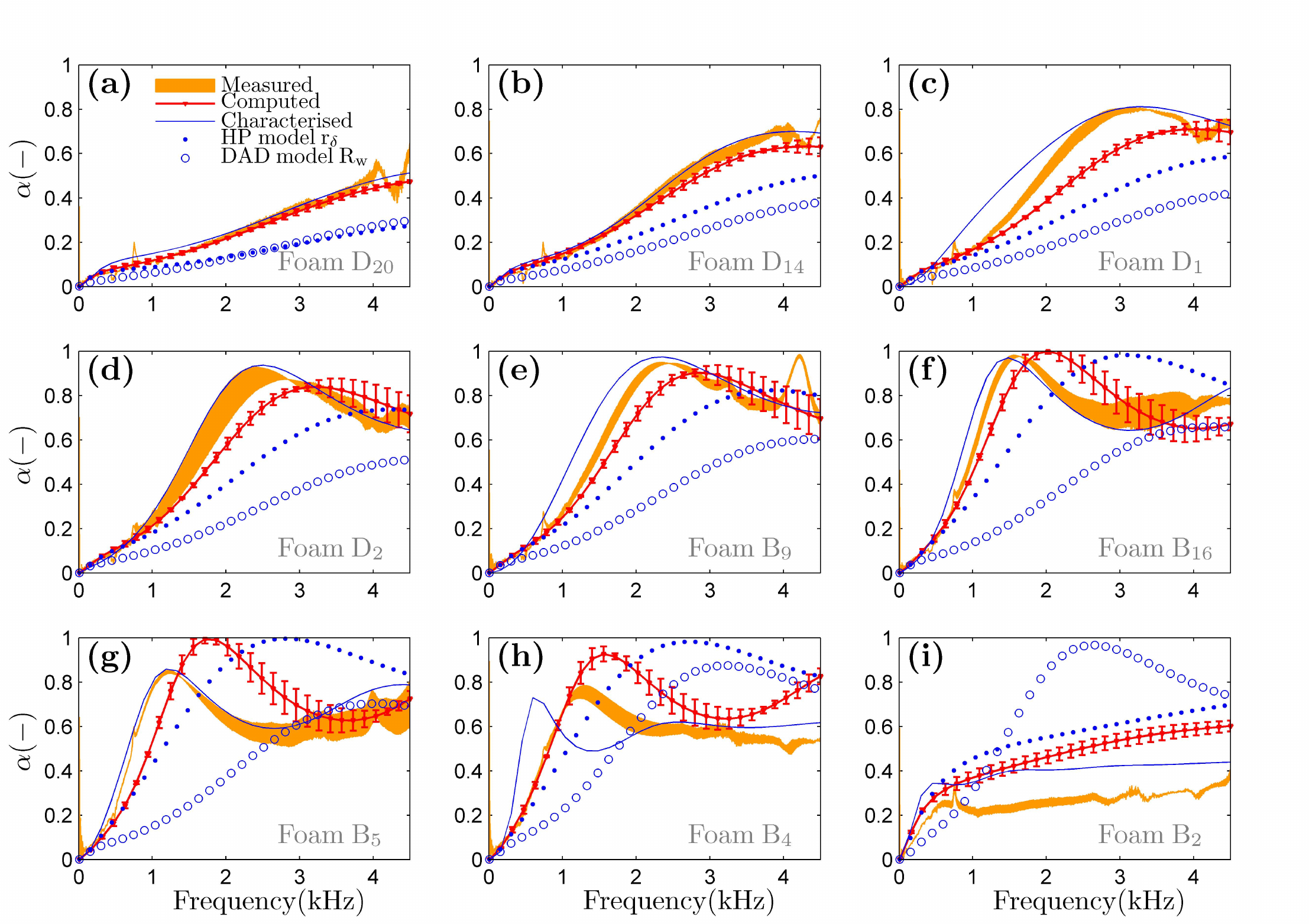}
\caption{\label{fig:FIG7}{(Color online) Sound absorption coefficients (SAC) of samples: experiments (filled zone), computations (thick continuous lines), characterizations (thin continuous lines), HP method (point markers), and DAD method (circle markers).}}
\end{figure}

\begin{figure}[!h]
\centering \includegraphics[width=1\textwidth, trim={0cm 0cm 1cm 0cm},clip]{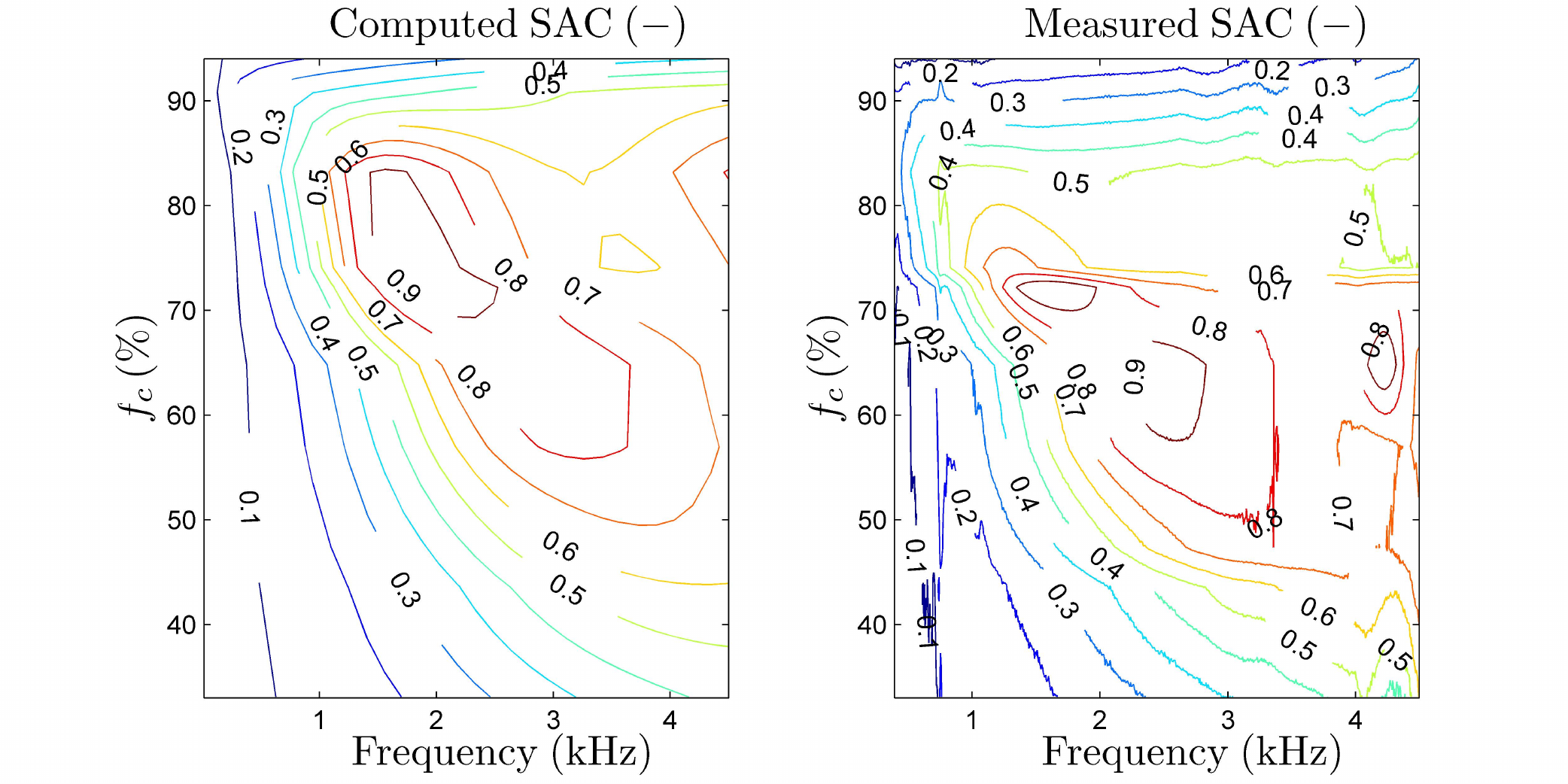}
\caption{\label{fig:FIG8}{(Color online) Illustration of SAC:  computations (left panel) and measurements (right panel).}}
\end{figure}

Our predictions are also compared with measurements and models available in the literature. The indirect method as proposed by Panneton and Only \cite{panneton2006, only2008} provides an experimental characterization of the macroscopic parameters ($\Lambda$, $\alpha_{\infty}$, $\Lambda$ and $k'_0$) from measured effective density and bulk modulus (assuming a 3-microphones impedance tube, and available estimates of $\sigma$ and $\phi$) by using analytical solutions derived from the JCAL model (see Appendix B). A simplified approach, referred to as a semi-empirical (2-parameter) model, has been proposed by Doutres et al. \cite{doutres2013} in assuming a micro-/macro correlation based on the cell size $D_b$ and a reticulation rate $R_w=x_p$, (see Appendix C). The third model is a numerical approach having the same procedure with our work but the different reconstructed PUCs including a membrane rate $r_\delta$. This provided by Hoang and Perrot \cite{hoang2012,hoang2014}, this work uses a PUC having membrane of equal size formed in all windows. This size increases from 0 (for open structure), then continuously increase to fully close squares and finally to cover the hexagons (closed cell structure). For a given porosity and cell size, following an increase of membrane size, the numerical computations are performed to investigate the varying of permeability. The PUC, having a membrane size that produces a similar numerical permeability in comparison with experimental or characterized data, is selected to represent foam material. (see Table S.V-VI in Supplemental Material for a detailed view about calculated transport parameters of the literature model of $r_{\delta}$).

Fig. \ref{fig:FIG7} shows that Doutres et al. (DAD) model (semi-empirical method) as well as Hoang and Perrot (HP) model (numerical method with equivalent membrane closure rate) fail to predict sound absorbing behavior of such foams which have both partially and fully closed windows. Note that these original models shown very good predictions of SAC in their studied materials with mostly partially closed windows (see SEM image of materials, Figure 1 in Refs. \cite{hoang2012,hoang2014} and Figure 8 in Ref. \cite{perrot2012}), or fully closed windows and opened windows (see Figure 1 in Ref. \cite{doutres2011}). How to use these models to characterize membrane foams? A detailed discussion on this statement will be described in the forthcoming part.

\subsection{Comment on models of acoustic membrane foams}
In a view point of modeling acoustical membrane foams, some interesting comments can be made based on the predicting ability of transport properties and SAC of the proposed as well as comparative models. Because of neglecting fully closed windows \cite{perrot2012, hoang2012,hoang2014} or partially closed windows \cite{doutres2013,doutres2011}, two existing models do not capture the wave propagation phenomena in advanced morphology of real foams composing of both fully and partially faces (with very poor prediction of the sound absorbing behavior, see Fig. \ref{fig:FIG7}). The presented model allows for accounting such materials and handles the limitations of previous models. We suggest that (i) the periodic unit cell for finite element simulation should involve a number of fully closed faces, and (ii) the semi-empirical approach with reticulated rate $R_w$ or cell openness $p$ should also including a fraction of partially closed faces ($x_p$) and corresponding their closure rate of membrane ($r_c$). Of course, by adding a number of fully closed faces in PUC used to numerical simulations, our computed results show a very good prediction in compared with results based HP estimates (see the points in Fig. \ref{fig:FIG7}) and Park's work (see Figure 6.(a) in Ref. \cite{park2017}). This is clearly confirmed for the first suggestion, and also for the second one as follows.

\begin{figure}[!h]
\centering \includegraphics[width=1\textwidth, trim={0cm 0cm 1cm 0cm},clip]{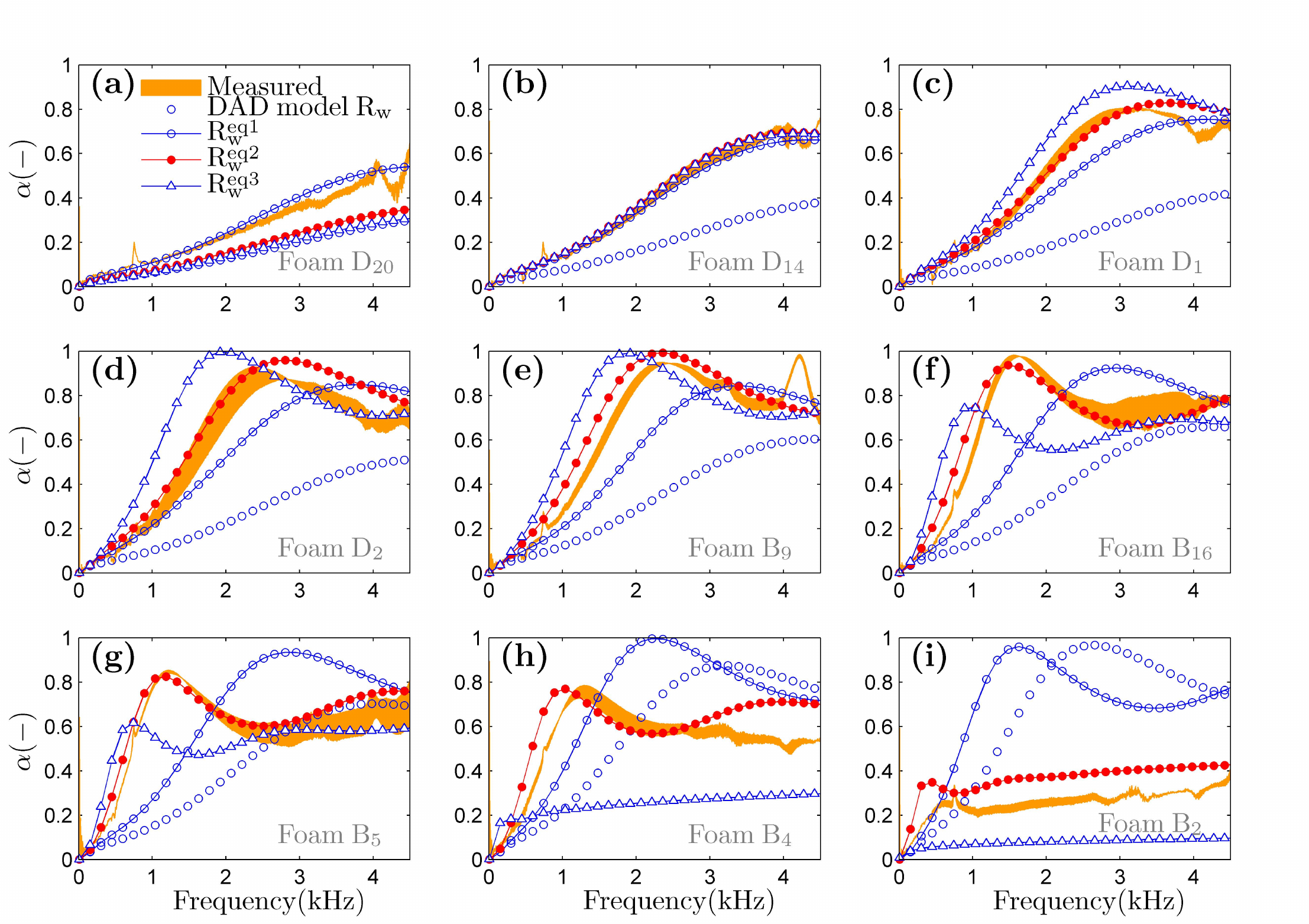}
\caption{\label{fig:FIG9}{(Color online) Sound absorption coefficients defined from the developed semi-empirical model with an equivalent reticulated rate: experiments (filled zone), computations (thick continuous lines), original DAD method (circle markers), developed DAD model (solid lines with markers: $R_{w}^{eq1}$ (circles), $R_{w}^{eq1}$ (filled circles), $R_{w}^{eq1}$ (filled triangles)).}}
\end{figure}

In the next, we develop the DAD semi-empirical estimates based on an equivalent reticulated rate $R_w^{eq}$ that characterized by different ways. The first equivalent reticulated rate is defined directly based on the morphology characterizations of local geometry of materials as,
\begin{equation}
R_w^{eq1}=x_p-r_cx_p
\end{equation}
The second equivalent reticulated rate is estimated by an equivalent macroscopic transport property for the case of airflow resistivity. It means that Eq. (\ref{eq:semi-resis1}) in Appendix C is used to find this equivalent ratio,
\begin{equation}\label{eq:comment-geo}
R_w^{eq2}=\Bigg (\frac{\sigma^{open}}{\sigma}\Bigg )^{{1}/{1.1166}}
\end{equation}
Similarly, considering a semi-empirical model with an equivalent high frequency tortuosity, the third equivalent reticulated rate is defined by,
\begin{equation}
R_w^{eq3}=\Bigg (\frac{\alpha^{open}_{\infty}}{\alpha_{\infty}}\Bigg )^{{1}/{0.3802}}
\end{equation}
in which, $\sigma^{open}=1674$ Nms$^{-4}$ and $\alpha_{\infty}^{open}=1.05$ and are airflow resistivity and high frequency tortuosity obtained from our computations for open cell structure foams (with cell size of $D_b$ and porosity $\phi$), respectively, resistivity; $\sigma$ and tortuosity $\alpha_{\infty}$ are characteristic values (Table S XX Supplemental Material). These results are very consistent with the given value $\alpha_{\infty}^{open}=1.05$ in Eq. (\ref{eq:semi-alpha1}) and the analytical result $\sigma^{open}=C^{\beta}(C_r^{\rho}r/L^2)^2=1781$ Nms$^{-4}$ in Eq. (\ref{eq:semi-resis1}) 

As shown in Fig. \ref{fig:FIG9}, it is seen clearly that all prediction curves of SAC behavior (lines with markers) has a fitting improvement by three equivalent rates in comparison with that from the original rate of proportion of open windows (circle markers). Here, we just employ new equivalent reticulated rate, and all of equations in the semi-empirical model has no modifying as shown in Appendix C. In particular, semi-empirical model with the reticulated rate $R_w^{eq1}$ shows a good SAC predictions only for two samples D20 and D14 with a lower membrane levels (see the thin lines with circular marker in Fig. \ref{fig:FIG8}(a-b)), and the model of $R_w^{eq3}$ shows a quite good predictions of SAC for samples named D14, D1, D2 and B9 (see Fig. \ref{fig:FIG9}(b-e), the continuous lines with filled triangular marker). In contrary, for the rate $R_w^{eq2}$ deduced from equivalent of resistivity, it is seen that this method can accurately model the sound absorption behavior of the closed cell foam materials in whole range of membrane level (see Fig. \ref{fig:FIG9}(b-i)). Interestingly, the comparison between two SAC results obtained from characterized work (the thin continuous line in Fig. \ref{fig:FIG7}) and semi-empirical model of equivalent rate $R_w^{eq2}$ (the continuous line with filled circular marker in Fig. \ref{fig:FIG9}) shows a very good agreement. It can be concluded that the empirical model developed based on equivalent tortuosity $\alpha_{\infty}$ or resistivity  $\sigma$ can predict accurately the sound absorption capability. The transport properties of all semi-empirical models are listed in Table S.VII in Supplemental Material.

Let us back with the above the HP numerical model. Even though it is also proposed based on the equivalent viscous permeability, this model fails to simulate the SAC of studied materials (for both cases identical membrane size ($r_{\delta}$) and identical membrane closure rate ($r^1_{\delta}$), see Fig .S.8 in Supplemental Material). Our model can capture sound absorbing properties, although our numerical and experimental permeability have a slightly difference. This means that the task of reconstruction of PUC plays a significant role in the numerically simulations of such membrane foam materials, because this requires a capturing the influence of not only permeability but also other macroscopic transport properties of materials.
\section{\label{sec:conclusion} Conclusion}
A three-dimensional idealized periodic unit cell (PUC)-based method to obtain the acoustic properties of high porous foam samples was described in this work. The PUC based on regular truncated octahedron packing is used for this hybrid multi-scale modeling. In this approach, first, some laboratory measurements of porosity and image processing techniques are taken. Different unit cells are then reconstructed from bubble size and membrane morphology characterization to compute macroscopic parameters from numerical homogenization. These later values serve in a sense as bridges between microstructure and acoustical macro-behavior with microphysical and micromechanical foundations. The numerical results are generally in good agreement with experimental values obtained from the measurements of standing wave tube.

As illustration of the obtained sound absorbing chart, it can be seen clearly that for a certain foam within given porosity and pore size, we are able to control their membrane morphology to archive the target sound absorption capacity in the frequency range of interest. For 20-mm thick layer made from membrane cellular foams with cell size 0.8 mm and porosity of 0.98, to reach the high absorption quality, the membrane level is should be kept in the range of 45 to 85$\%$ by using the gelatin concentration $C_{gel}$ around 16$\%$. In this target range of membrane fraction, it can be noted that the foam absorber can archive the high SAC in the high frequency band with $f_c$ around value of lower bound of 80$\%$, and the low frequency band with $f_c$ round the value of upper bound of 50$\%$.
The context of foaming morphology in our proposed approach can be adopted for other purpose in mechanical investigation of such foam-based materials. The development of an advanced modeling of the membrane content as well as few comments for literature models, such as the one proposed in this communication.

\begin{acknowledgments}
This work was part of a project supported by ANRT (Grant no. ANR-13-RMNP-0003-01). The work of V. H. Trinh was supported by a fellowship awarded by the Government of Vietnam (Project 911).
 \nameref{subsec:AppendixA}
\end{acknowledgments}
\begin{appendix}
\section*{\label{sec:Appendix} Appendix}
\subsection*{Appendix A: Reconstruction of  periodic unit cell}
\label{subsec:AppendixA}
\setcounter{equation}{0}
\renewcommand\theequation{A.\arabic{equation}}
In this appendix, we reconstruct the periodic unit cell based on Kelvin pattern. A part of the 1/96 periodic unit cell is approximated as 1/4 triangular tube and 1/8 octahedron placed at their junction of node (see Fig. \ref{fig:FIG10}). It is easy to find the coordinates of 7 vertices of this skeleton listed as follows: A$(0,r/2\text{tan}{\beta},0)$; B$(r/2,0,0)$; C$( 0,-r/2\text{tan}{\beta},0)$; F$(r/2,0,-r/2)$; J$(r/2+L_l \sqrt 2/4,-L_l \sqrt2/4,0)$; K$(L_l \sqrt2/4,-L_s \sqrt2/4,0)$ and M$(r/2+L_l \sqrt2/4,-L_l \sqrt2/4,-r/2)$. In which, $L_l=(D_b/4-r/2)\sqrt 2$ ,$L_s=[D_b/4-r/2(\sqrt6-1)] \sqrt2$ , $\beta=3\pi/4-\alpha$ with tan$\alpha=\sqrt 3/(\sqrt3-\sqrt2)$.
\begin{figure}[!h]
\centering \includegraphics[width=1\textwidth, trim={0cm 0cm 0cm 0cm},clip]{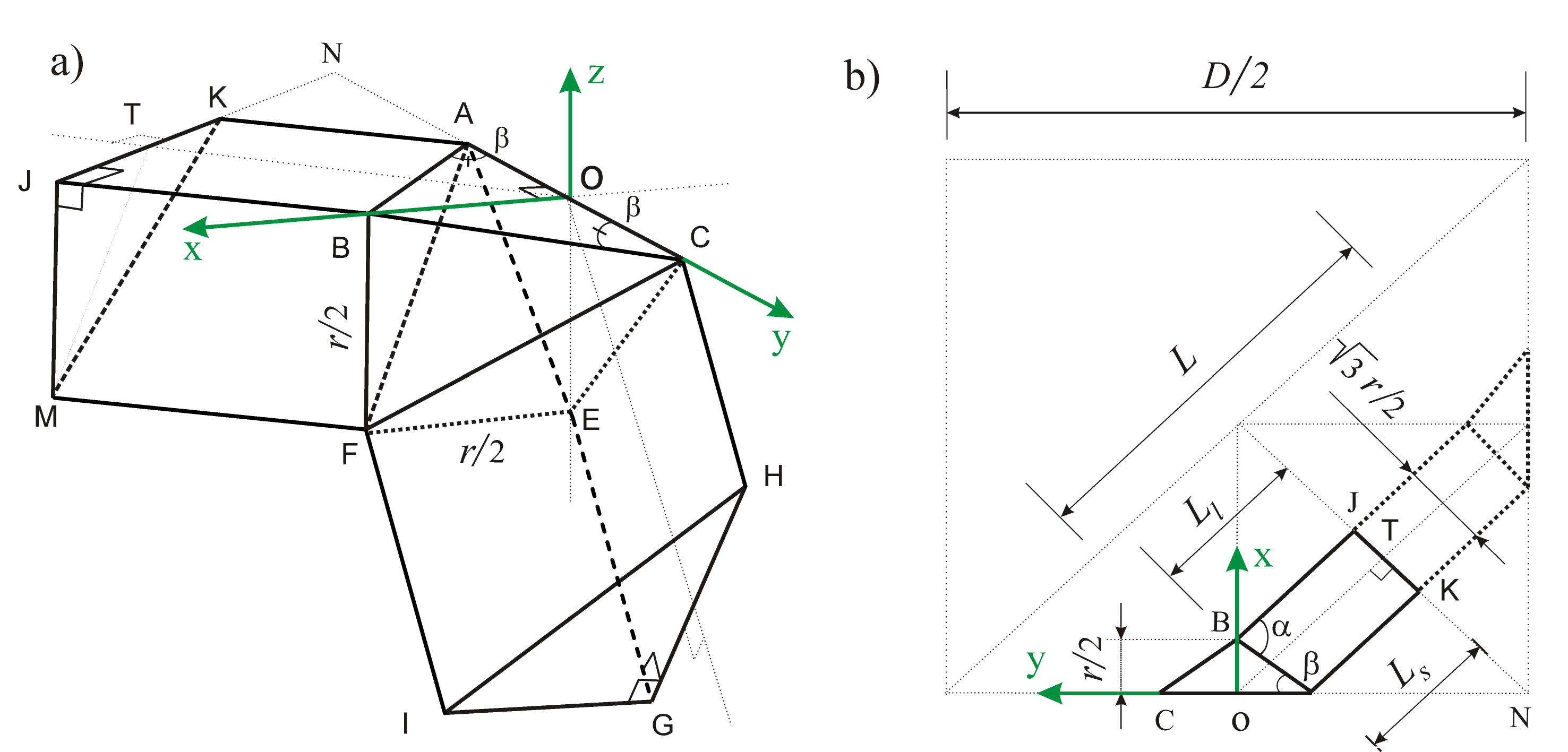}
\caption{\label{fig:FIG10}{(a) Detail of the coordinates of the basic vertex in a 1/48 open unit cell having ligaments of equilateral triangular cross section of edge size $r$. (b) Diagram shows the relations between the angular and length parameters of node and ligament (top view of the skeleton).}}
\end{figure}

The total solid volume of unit cell is the volume of 12 nodes and 24 ligaments (12 edges on hexagonal faces and 24 edges shared with the neighboring cells), given by
\begin{equation}
V_s=96\times(V_{\rm{ABFJKM}}+V_{\rm{ABCF}})=
6\sqrt 3 Lr^2+(4\text{tan}\beta-6-\sqrt 6)r^3
\end{equation}
The porosity of the open cell structure can be defined as,
\begin{equation}\label{eq:poro-open}
\phi=1-\frac{3\sqrt 6}{16}\Bigg(\frac{r}{L} \Bigg)^2-\frac{4\text{tan}\beta-6-\sqrt 6}{16\sqrt 2}\Bigg(\frac{r}{L} \Bigg)^3
\end{equation}
Solving Eq. (\ref{eq:poro-open}), it is obtained the ligament size $r/L$ as a function of porosity $\phi$,
\begin{equation}\label{eq:approx_radii}
\frac{r}{L}=\frac{P_2}{3P_3}\left[1-\text{cos}\frac{\text{arcos}(\chi)-2\pi}{3}\right]
\end{equation}
where $\chi=\frac{P_2^3-27P_3P_0}{2\sqrt{P_2^3}}$,
with $P_0=1-\phi$, $P_2=\frac{3\sqrt 6}{16}$, and $P_3=\frac{4\text{tan}\beta-6-\sqrt 6}{16\sqrt 2}$. Hence, one obtain an approximate expression from Eq. (\ref{eq:approx_radii}) as $r/L=0.5833(1-\phi)^{0.521}$.
\appendix\hypertarget{Appendix B}{\subsection*{Appendix B: Numerical estimations of transport property} }
\setcounter{equation}{0}
\renewcommand\theequation{B.\arabic{equation}}
Two purely geometrical parameters ($\phi$, $\Lambda'$) are defined directly from the local geometry via an unit cell $\mathrm\Omega$ (of $\mathrm\Omega_f$ fluid-filled domain and $\partial\mathrm\Omega$ fluid-solid interface) as,
\begin{equation}
 \phi=\frac{\int_\mathrm{\Omega_f}dV}{\int_\mathrm{\Omega}dV},\ \ \ \ \
 \\ \Lambda^\prime=2\frac{\int_\mathrm{\Omega}dV}{\int_{\partial\mathrm{\Omega}}dS}
 \end{equation}
The remaining transport property is computed from the numerical solutions of three group of governing equations into the unit cell. Firstly, the low Reynolds number flow of an incompressible Newtonian fluid is governed by the usual Stokes equations in the fluid phase $\mathrm\Omega_f$,\cite{auriault2009}:
\begin{subequations}
\begin{equation}
\eta\mathbf{\Delta v }- \nabla p =  - G\ \text{with}\
 \ \nabla. \textbf{v} = 0 \ \  \text{in}\  \  \mathrm\Omega_f
\end{equation}
\begin{equation}
\textbf{v}=0\ \  \text{on} \ \partial\mathrm\Omega
\end{equation}
\begin{equation}
\textbf{v}\ \text{and}\  {p}\ \  \text{are}\ \mathrm{\Omega} - \text{periodic}
\end{equation}
\end{subequations}
where $G=\nabla p^m$ is the macroscopic pressure gradient acting as a source term. Symbols $\mathbf{v}$ and $p$ are the velocity and pressure of the fluid respectively. In general, $\mathbf v$ satisfies the non-slip condition ($\mathbf{v } $=0) at $\partial{\mathrm\Omega}$. It can be shown that the local field of the static viscous permeability are  obtained from the local velocity field as,
\begin{equation}
\textbf{k}_0=-\frac{\eta}{G}\textbf{v}
\end{equation}
The static viscous permeability $k_0$ and static viscous tortousity $\alpha_0$ are calculated by the standard definitions below,
\begin{equation}
k_0=\phi\left\langle \textbf{k}_0\right\rangle,\ \ \ \ \
\alpha_0=\frac{\left\langle \textbf{k}^2_0\right\rangle}{\left\langle \textbf{k}_0\right\rangle^2}
\end{equation}
The symbol $\langle {\textbf{.}}\rangle$ denotes a fluid-phase averaging operator, $\langle {\textbf{.}}\rangle=\int_{\mathrm\Omega_f}(\textbf{.}){dV}$.

Secondly, at the high frequency range with $\omega$ large enough, the viscous boundary layer becomes negligible and the fluid tends to behave as a perfect one, having no viscosity except in a boundary layer. Consequently, the perfect
incompressible fluid formally behaves according to the electrical conduction problem \cite{achdou1992,avellaneda1991,perrot2008}:
\begin{subequations}
\begin{equation}
 \nabla\textbf{.E}=0 \ \ \text{with}\ \
 \textbf{E}=-\nabla\varphi+\textbf{e},\ \ \ \text{in}\ \ {\mathrm\Omega_f}
\end{equation}
\begin{equation}
\textbf{E.n}=0, \ \ \ \text{in}\ \ \partial{\mathrm\Omega_f}
\end{equation}
\begin{equation}
\varphi\  \text{is}\ \mathrm{\Omega} - \text{periodic}
\end{equation}
\end{subequations}
where $\textbf{e}$ is a given macroscopic electric field, $\textbf{E}$ the solution
of the boundary problem having $-\nabla\varphi$ as a fluctuating part, and $\textbf{n}$ is unit normal to the boundary of the pore region.
\\
The high frequency tortuosity $\alpha_{\infty}$ and the viscous characteristic length $\Lambda$ are calculated:
\begin{equation}
\alpha_{\infty}=\langle\textbf{E.e}\rangle,\ \
\Lambda=\frac{\langle\textbf{E}^2\rangle}{\langle\textbf{E}\rangle^2}
\end{equation}
Now, all transport parameters of reference model are available.
\appendix\hypertarget{Appendix C}{\subsection*{Appendix C: Semi-empirical estimations of transport property} }
\setcounter{equation}{0}
\renewcommand\theequation{C.\arabic{equation}}
This estimation is provided from reference model by Doutres et al.\cite{doutres2013,doutres2011} for membrane cellular foams. In this model, a simplified micro-/macro link is presented based on a cell size $D_b$ and a reticulated rate $R_w$. This work uses the classical JCA model to predict the sound absorption efficiency, this model is known as the 5-parameters ( $\phi$, $\Lambda'$, $\Lambda$,  $\sigma$, and $\alpha_{\infty}$).
Firstly, the porosity and the thermal characteristic length of materials is calculated as,
\begin{equation}
\phi=\frac{C_p}{B^2},\ \Lambda'=D_b\frac{8\left[1-(2\sqrt3-\pi)/B^2\sqrt2 \right]/3A}{1+2\sqrt 3-R_w(1+2\sqrt 3-4\pi/B\sqrt3)}
\end{equation}
where $C_p=(2\sqrt3-\pi)/\sqrt2$, $A=D_b/\sqrt2$ and $B=D/(Ar\sqrt 2)$.

Then, the viscous characteristic length is defined as a function of the reticulated rate $R_w$,
\begin{equation}
\Lambda'/ \Lambda=1.55/R_w^{0.6763}
\end{equation}
The high frequency tortuosity is defined as,
\begin{equation}\label{eq:semi-alpha1}
\alpha_{\infty}=1.05/R_w^{0.6763}
\end{equation}
Finally, the airflow resistivity is deduced by,
\begin{equation}\label{eq:semi-resis1}
\sigma=C^{\beta}(C_r^{\rho}r/L^2)^2/R_w^{1.1166}
\end{equation}
in which $C^{\beta}=128\alpha_{\infty}\eta/c^2_g$ and $C_r^{\rho}=3\pi/8\sqrt 2$.
\appendix\hypertarget{Appendix D}{\subsection*{Appendix D: Characterized estimations of transport and effective property} }
\setcounter{equation}{0}
\renewcommand\theequation{D.\arabic{equation}}
The characterization method \cite{panneton2006,only2008} is presented for determining the normal incidence sound absorption coefficient and the effective properties of tested materials based on the experimental data with two measured pressure transfer functions $H_{12}$ and $H_{23}$.
The complex reflection coefficient of sample is deduced:
\begin{equation}
R=\frac{\text{exp}(jd_{12}k_0)-H_{12}}{H_{12}-\text{exp}(-jd_{12}k_0)}\text{exp}(2jk_0L_s)
\end{equation}
where $d_{12}$ is the distance between microphone 1 and 2 (see Fig. 3), $k_0$ denotes the wave number in the ambient fluid.

The pressure ratio $H_0$ between the front and the rear of the sample is estimated as,
\begin{equation}
H_0=\frac{1+R}{\text{exp}(jk_0L_s)-R/\text{exp}(jk_0L_s)}H_{23}
\end{equation}
Then, the wave number $k_c(\omega)$ and the characteristic impedance $Z_c(\omega)$ are given by,
\begin{equation}
k_c(\omega)=\frac{1}{L_s \text{cos}(H_0)}, \  Z_c(\omega)=jZ_s\text{cot}(k_c(\omega)L_s),~\text{with}~Z_s/Z_0=\frac{1+R}{1-R}
\end{equation}
The effective density and the effective bulk modulus properties of materials are also evaluated as,
\begin{equation}
\rho(\omega)= Z_c(\omega)k_c(\omega)/\omega, \  K(\omega)=Z_c(\omega)\omega/k_c(\omega)
\end{equation}
Finally, five macroscopic properties of materials are determined from the measured effective density $\rho(\omega)$ and bulk modulus $K(\omega)$ using the indirect method proposed by Panneton and Olny[38, 39] as follows,
\begin{equation}\label{eq:charac-resis}
\sigma=\lim_{\omega\to 0}\left[ \Im(\omega \rho) \right]
\end{equation}
\begin{equation}
\alpha_{\infty}=\frac{\phi}{\rho_0}\left[ \Re (\rho)-\sqrt{\Im (\rho)^2-\Bigg(\frac{\sigma \phi}{\omega}\Bigg)^2} \right]
\end{equation}
\begin{equation}
\Lambda=\frac{\alpha_{\infty}}{\phi}\sqrt{\frac{2\rho_0\eta}{\omega\Im(\rho)(\rho_0\alpha_{\infty}/\phi-\Re(\rho))}}
\end{equation}
\begin{equation}
\Lambda'=2\sqrt{\frac{\eta}{P_r\rho_0\omega}\left\{-\Im\Bigg(\Bigg[\frac{1-K/K_a}{1-\gamma K/K_a}\Bigg]^2\Bigg)
 \right\}}
\end{equation}

\begin{equation}
k_0'=\frac{\phi\eta}{P_r\rho_0\omega}\left\{-\Re\Bigg(\Bigg[\frac{1-K/K_a}{1-\gamma K/K_a}\Bigg]^2\Bigg)
 \right\}^{-1/2}
\end{equation}
in which $K_a$ denotes adiabatic bulk modulus.
\end{appendix}




%
%
%
%
%
%
%





\end{space}

\end{document}